\documentstyle[amstex,amssymb,12pt,epsf]{article}

\begin{document}

\begin{center}
{\bf {\Large
Time-reversal-violating birefringence of photon in a 
medium exposed to electric and magnetic field.
}}

\bigskip

Vladimir G.Baryshevsky

Nuclear Problem Research Institute,

Bobruiskaya Str.11, Minsk 220080 Belarus.

Electronic address: bar@@inp.minsk.by

Tel: 00375-172-208481, Fax: 00375-172-265124

\bigskip
\end{center}

\section{Introduction}

Violation of the time reversal symmetry has been observed in $K_{0}$-decay only 
\cite{1,2_new}, and remains one of the greatest unsolved problems in the elementary
particle physics. A lot of attempts have been undertaken to observe this
phenomenon experimentally in different processes. However, those experiments
have not been successful.

At the present time novel more precise experiments are actively discussed.
Let us note here the experiments with atom (molecule) and neutron spin
precession in an electric field $\overrightarrow{E}$\ due to interaction of
the electric dipole moment (EDM) of atom (molecule) or neutron with the
field \ $\overrightarrow{E}$ \cite{2,3}. The experiments are discussing for 
observation of the polarization plane rotation phenomena caused by the pseudo-Zeeman
splitting of the atom(molecule) levels by an external electric field \ $%
\overrightarrow{E}$\ due to interaction $W=-\overrightarrow{d_{a}}\cdot 
\overrightarrow{E}$\ of the atom (molecule) EDM $\overrightarrow{d_{a}}$
with an electric field [5-9](this effect is
similar to the magneto-optic Macaluso-Corbino effect  \cite{Macaluso}).

It should be noted that the mentioned experiments use the possible existence
of such intrinsic quantum characteristic of an atom (molecule) as a static EDM.
According to \cite{4,5,6} together with the EDM there is one more
characteristic of atom (molecule) describing its response to the external
field effect - the T- and P-odd polarizability of atom (molecule) $\beta
^{T} $. This polarizability differs from zero even if the EDM of an electron is
equal to zero and pseudo-Zeeman splitting of atom (molecule) levels is
absent.

\bigskip Both $\beta ^{T}$ and EDM yield to appearance of several new optical
phenomena. Let us mention two of them:

a). the T-odd photon birefringence effect \cite{14_new} (i.e effect when plane polarized
photons are converted to circular polarized ones and vice versa, this effect
is similar to magneto-optic birefringence{\large \ }Cotton-Mouton effect 
\cite{Landau8})

b). the photon polarization plane rotation and circular dichroism in an
optically homogeneous isotropic medium exposed to an electric field caused
by the Stark mixing of atom (molecule) levels \cite{lanl99,PLA}. This T-odd
phenomenon is a kinematic analog of the well known T-even phenomenon of \
Faraday effect of the photon polarization plane rotation in the medium
exposed to a magnetic field due to Van-Vleck mechanism. Similarly \ to the
well known P-odd T-even \ effect of light polarization plane rotation for
which the intrinsic spin spiral of atom is responsible \cite{7}, this effect
is caused by the atom magnetization appearing under external electric field
action. \ Moreover, according to \cite{lanl99}, the magnetization of atom
appearing under action of static electric field causes the appearance of
induced magnetic field \ $\overrightarrow{H_{ind}}(\overrightarrow{E})$ .
The energy of interaction of \ atom magnetic moment $\overrightarrow{\mu _{a}%
}$\ with this field is \ $W_{H}=-\overrightarrow{\mu _{a}}\cdot $\ $%
\overrightarrow{H_{ind}}(\overrightarrow{E}).$\ Therefore, the total
splitting of atom levels is determined by energy {\em \ }$W_{T}=-%
\overrightarrow{d_{a}}\cdot \overrightarrow{E}-\overrightarrow{\mu _{a}}%
\cdot \overrightarrow{H_{ind}}(\overrightarrow{E})$. As a result, the effect
of polarization plane rotation (birefringence effect) deal with the energy
levels splitting is caused not only by $\overrightarrow{d_{a}}$ interaction
with electric field\ but by $\overrightarrow{H_{ind}}(\overrightarrow{E})$
interaction with $\overrightarrow{\mu _{a}}$, too.\ It is easy to see, that
even for $\overrightarrow{d_{a}}=0$\ the energy of splitting differs from
zero and the T-odd effect of polarization plane rotation and birefringence
exist.

In that way the time reversal violating weak interactions yield to the
appearance of diverse interesting optical phenomena. In the present
paper the optical anisotropy of medium caused by T-odd interactions of
electrons and nuclei in atoms and molecules is investigated. The expressions
for T-odd polarizabilities of atoms (molecules) in external electric and
magnetic fields are obtained. Magnitude of light polarization plane rotation
and birefringence effects are estimated. It is shown that investigation of
optical anisotropy caused by T-odd interactions can provide\ information
about constants of T-odd weak interactions.

\section{\protect\Large Tensor of dielectric permittivity of medium at
presence of T, P-odd weak interactions.}

Let us consider the propagation of electromagnetic wave through the
homogeneous isotropic medium placed to the stationary electric\ $%
\overrightarrow{E}$\ and magnetic \ $\overrightarrow{H}$\ fields. Tensor of
dielectric permittivity of medium $\varepsilon _{ik}$ is a certain function
of these fields

\[
\varepsilon _{ik}=\varepsilon _{ik}(\omega ,\overrightarrow{k},%
\overrightarrow{E},\overrightarrow{H}), 
\]

where $\omega $ is the frequency and $\overrightarrow{k}$\ is the wave
vector of the photon.

Tensor $\varepsilon _{ik}$ can be presented as the sum

\[
\varepsilon _{ik}=\varepsilon _{ik}^{even}(\omega ,\overrightarrow{k},%
\overrightarrow{E},\overrightarrow{H})+\varepsilon _{ik}^{odd}(\omega ,%
\overrightarrow{k},\overrightarrow{E},\overrightarrow{H}), 
\]

where $\varepsilon _{ik}^{even}(\omega ,\overrightarrow{k},\overrightarrow{E}%
,\overrightarrow{H})$\ \ is the tensor of dielectric permittivity of medium
in the absence of P,T-odd interactions and \ $\varepsilon _{ik}^{odd}(\omega
,\overrightarrow{k},\overrightarrow{E},\overrightarrow{H})$ is the term
caused by the T,P-odd weak interactions. Let us expand $\varepsilon _{ik}$
into irreducible parts:

\begin{equation}
\varepsilon _{ik}=\varepsilon _{0}\delta _{ik}+\varepsilon
_{ik}^{s}+\varepsilon _{ik}^{a},  \label{expand}
\end{equation}

where $\delta _{ik}$\ is the Kronecker symbol, indices $i,k=1,2,3$
correspond (x,y,z), $\varepsilon _{0}=\frac{1}{3}\sum_{i}\varepsilon _{ii}$
is the scalar, $\varepsilon _{ik}^{s}=\frac{1}{2}(\varepsilon
_{ik}+\varepsilon _{ki})-\varepsilon _{0}\delta _{ik}$ is the symmetric
tensor (with trace equal to zero), $\varepsilon _{ik}^{a}=\frac{1}{2}%
(\varepsilon _{ik}-\varepsilon _{ki})$ is the antisymmetric tensor of rank
two.

\bigskip  An antisymmetric tensor of rank two can be
represented (see, for example \cite{14_new} \ as

\[
\varepsilon _{ik}^{a}=ie_{ikl}g_{l,} 
\]

where $\overrightarrow{g}$\ is the axled vector dual to the antisymmetric
tensor $\varepsilon _{ik}^{a},$ $e_{ikl}$ is the fully antisymmetrical unit
tensor of rank three.

\bigskip Tensors $\varepsilon _{ik}^{even}(\omega ,\overrightarrow{k},%
\overrightarrow{E},\overrightarrow{H})$ and $\varepsilon _{ik}^{odd}(\omega ,%
\overrightarrow{k},\overrightarrow{E},\overrightarrow{H})$ can be written by
way of (\ref{expand}), too.

\bigskip

Tensor $\varepsilon _{ik}^{even}(\omega ,\overrightarrow{k},\overrightarrow{E%
},\overrightarrow{H})$ describes optical anysotropy of the media in external
fields for P,T-even world. Particularly, the symmetrical part of $%
\varepsilon _{ik}^{even}$ is responsible for birefringence effects caused by
the external fields $\overrightarrow{E}$ and $\overrightarrow{H}$ (i.e.
effects of Kerr and Cotton-Mouton) and antisymmetrical part describes
Faraday and Macaluso-Corbino effects.

\bigskip

Let us consider $\varepsilon _{ik}^{odd}(\omega ,\overrightarrow{k},%
\overrightarrow{E},\overrightarrow{H})$. Weak interactions are lower than
electromagnetic ones. Then, only linear terms\ in the expansion of $%
\varepsilon _{ik}^{odd}$ over weak interaction constant should be taken into
consideration. Therefore, $\varepsilon _{ik}^{odd}$ depends on P,T-odd
scalar and tensor combinations of vectors $\overrightarrow{E},%
\overrightarrow{H}$ and $\overrightarrow{k}$ linearly:

\begin{eqnarray}
\varepsilon _{ik}^{odd} &=&\varepsilon _{0}^{odd}\delta _{ik}+\varepsilon
_{ik}^{(s)\;odd}+ie_{ikl}g_{l}^{odd}=  \label{e_odd} \\
&=&[\chi _{sEH}^{T}(\overrightarrow{n_{E}}\;\overrightarrow{n_{H}})+\chi
_{sE}^{T}(\overrightarrow{n_{\gamma }}\;\overrightarrow{n_{E}})+\chi
_{sH}^{P}(\overrightarrow{n_{\gamma }}\;\overrightarrow{n_{H}})]\delta _{ik}+
\nonumber \\
&&+\chi _{t}^{T}[\frac{1}{2}(n_{E\,i\;}n_{H\,k\;}+n_{H\,i\;}n_{E\,k\;})-%
\frac{1}{3}(\overrightarrow{n_{E}}\;\overrightarrow{n_{H}})]+  \nonumber \\
&&+i\chi _{sEH}^{P}e_{ikl}n_{EHl}+i\chi _{s}^{P}e_{ikl}n_{\gamma l}+i\chi
_{E}^{T}e_{ikl}n_{El}  \nonumber
\end{eqnarray}
where $\overrightarrow{n}_{\gamma }=\frac{\overrightarrow{k}}{\left| 
\overrightarrow{k}\right| }$,$\overrightarrow{n}_{E}=\frac{\overrightarrow{E}%
}{\left| \overrightarrow{E}\right| }$, $\overrightarrow{n}_{H}=\frac{%
\overrightarrow{H}}{\left| \overrightarrow{H}\right| }$ and $\overrightarrow{%
n}_{EH}=\frac{[\overrightarrow{E}\times \overrightarrow{H}]}{\left| [%
\overrightarrow{E}\times \overrightarrow{H}]\right| }$ are the unit vectors,
repeated indices imply summation.
Term, containing $\delta _{ik}$,\ describes contribution of P,T-odd weak
interactions in the scalar part of dielectric polarisability, where $\chi
_{sEH}^{T}$ determines contribution of T,P-odd interactions, $\chi _{sE}^{T}$
is responsible for T-odd, P-even interactions (this contribution was first
considered in \cite{19_new}), $\chi _{sH}^{P}$ describes P-odd, T-even
interactions. The second term in (\ref{e_odd}),\ which is proportional to $%
\chi _{t}^{T}$,\ is responsible for T,P-odd birefringence effect (predicted
in \cite{14_new}). Terms proportional to tensor $e_{ikl}$ describes
light polarisation plane rotation around $\overrightarrow{n}_{EH}$ (P-odd,
T-even), around electric field direction $\overrightarrow{n}_{E}$ (P,
T-odd) and around $\overrightarrow{n}_{\gamma }$ (P-odd, T-even). The latter is
well known and actively studied phenomenon of polarisation plane rotation
and dichroism caused by P-odd T-even interactions \cite{7,10,11}. If spins of 
atoms (molecules) are certainly oriented then additional
contributions to $\varepsilon _{ik}^{odd}$ can appear \cite{5}.

For further consideration let us suppose that medium is optically deluted ( $%
\varepsilon _{ik}-\delta _{ik}\ll 1$). Then the dielectric permittivity
tensor $\varepsilon _{ik}$ depends on the amplitude of elastic coherent
scattering of photon by atom (molecule) \cite{4,5,6}:

\begin{equation}
\varepsilon _{ik}=\delta _{ik}+\chi _{ik}=\delta _{ik}+\frac{4\pi \rho }{%
k^{2}}f_{ik}(0),
\end{equation}
where $\chi _{ik}$\ is the polarizability tensor of a medium, $\rho $\ is
the number of atoms (molecules) per $cm^{3}$, $k$\ is the photon wave
number; $f_{ik}(0)$\ is the tensor part of the zero angle amplitude of
elastic coherent scattering of a photon by an atom (molecule). Indices $%
i=1,2,3$ are referred to coordinates $x,y,z$, respectively.

Similarly $\varepsilon _{ik}$ tensor $f_{ik}(0)$ can be expanded in
irreducible parts:

\begin{eqnarray}
f_{ik}(0) &=&f_{ik}^{even}+f_{ik}^{odd}=f_{ik}^{even}+\frac{\omega ^{2}}{%
c^{2}}\alpha _{ik}^{odd}=  \nonumber \\
&=&f_{ik}^{even}+\frac{\omega ^{2}}{c^{2}}\left\{ [\beta _{sEH}^{T}(%
\overrightarrow{n_{E}}\;\overrightarrow{n_{H}})+\beta _{sE}^{T}(%
\overrightarrow{n_{\gamma }}\;\overrightarrow{n_{E}})+\beta _{sH}^{P}(%
\overrightarrow{n_{\gamma }}\;\overrightarrow{n_{H}})]\delta _{ik}\right. +
\label{amplitude_ik} \\
&&+\left. \beta _{t}^{T}[\frac{1}{2}(n_{E\,i\;}n_{H\,k\;}+n_{H\,i\;}n_{E\,k%
\;})-\frac{1}{3}(\overrightarrow{n_{E}}\;\overrightarrow{n_{H}})]\right. + 
\nonumber \\
&&+\left. i\beta _{sEH}^{P}e_{ikl}n_{EHl}+i\beta _{s}^{P}e_{ikl}n_{\gamma
l}+i\beta _{E}^{T}e_{ikl}n_{El}\right\}  \nonumber
\end{eqnarray}
where $\alpha _{ik}^{odd}$ is the tensor of dynamical polarizability of an
atom (molecule) and quantities $\beta$ are the dinamical polarizabilities of an
atom (molecule).

The amplitude of elastic coherent scattering of photon by atom (molecule) at
zero angle can be written as

\[
f(0)=f_{ik}(0)e_{i}^{\prime \ast }e_{k} 
\]

Here $\overrightarrow{e}$\ and $\overrightarrow{e}^{\prime }$ \ are the unit
polarization vectors of initial and scattered photons. The unit vectors
describing the circular polarization of photons are: $\overrightarrow{e}%
_{+}=-\frac{\overrightarrow{e}_{1}+i\overrightarrow{e}_{2}}{\sqrt{2}}\ $for
the right and, $\overrightarrow{e}_{-}=\frac{\overrightarrow{e}_{1}-i%
\overrightarrow{e}_{2}}{\sqrt{2}}$\ for the left circular polarization,
where $\overrightarrow{e}_{1}\perp \overrightarrow{e}_{2}\ $, $%
\overrightarrow{e}_{2}=[\overrightarrow{n}_{\gamma }\times \overrightarrow{e}%
_{1}]\ $are the unit polarization vectors of a linearly polarized photon, $[%
\overrightarrow{e}_{1}\times \overrightarrow{e}_{2}]=\overrightarrow{n}%
_{\gamma }$, $\overrightarrow{e}_{1}=-\frac{\overrightarrow{e}_{+}-%
\overrightarrow{e}_{-}}{\sqrt{2}}$, $\overrightarrow{e}_{2}=-\frac{%
\overrightarrow{e}_{+}+\overrightarrow{e}_{-}}{i\sqrt{2}}$.

The refractive index\ is as follows:

\begin{equation}
\widehat{N}=N_{ik}=1+\frac{2\pi \rho }{k^{2}}f_{ik}.
\end{equation}

Suppose $\overrightarrow{H}=0$\ and an electromagnetic wave propagates
through a gas along the electric field\ $\overrightarrow{E}$ direction. The
refractive indices for the right $N_{+}$\ and for the left $N_{-}$\ circular
polarized photons can be written as:

\begin{equation}
N_{\pm }=1+\frac{2\pi \rho }{k^{2}}f_{\pm }(0)=1+\frac{2\pi \rho }{k^{2}}%
\left\{ f^{ev}(0)+\frac{\omega ^{2}}{c^{2}}\beta _{sE}^{T}(\overrightarrow{%
n_{\gamma }}\;\overrightarrow{n_{E}})\mp \frac{\omega ^{2}}{c^{2}}\left[
\beta _{s}^{P}+\beta _{E}^{T}(\overrightarrow{n}_{E}\overrightarrow{n}%
_{\gamma })\right] \right\} ,
\end{equation}
where $f_{+}(0)(f_{-}(0))$ is the zero angle amplitude of the elastic
coherent scattering of the right (left) circular polarized photon by an atom
(molecule).

Let photons with the linear polarization $\overrightarrow{e}_{1}=-\frac{%
\overrightarrow{e}_{+}-\overrightarrow{e_{-}}}{\sqrt{2}}$\ fall in a gas.
The polarization vector of a photon in a gas $\overrightarrow{e}_{1}^{\prime
}$\ can be written as: 
\begin{eqnarray}
\overrightarrow{e}_{1}^{\prime }\ &=&-\frac{\overrightarrow{e}_{+}\ }{\sqrt{2%
}}e^{ikN_{+}L}+\frac{\overrightarrow{e}_{-}\ }{\sqrt{2}}e^{ikN_{-}L}=
\label{5} \\
&=&e^{\frac{1}{2}ik(N_{+}+N_{-})L}\left\{ \overrightarrow{e}_{1}\cos \frac{1%
}{2}k(N_{+}-N_{-})L-\overrightarrow{e}_{2}\sin \frac{1}{2}%
k(N_{+}-N_{-})L\right\} ,  \nonumber
\end{eqnarray}
where $L$\ is the photon propagation length in a medium.

As one can see, the photon polarization plane rotates in a gas. The angle of
rotation $\vartheta $\ is 
\begin{eqnarray}
\vartheta &=&\frac{1}{2}k {Re}(N_{+}-N_{-})L=\frac{\pi \rho }{k} {Re}\left[
f_{+}(0)-f_{-}(0)\right] L=  \label{6} \\
&=&-\frac{2\pi \rho \omega }{c}Re\left[ \beta _{s}^{P}+\beta _{E}^{T}(%
\overrightarrow{n}_{E}\overrightarrow{n}_{\gamma })\right] L,  \nonumber
\end{eqnarray}
where ${Re}N_{\pm }$\ is the real part of $N_{\pm }$. It should be noted
that $\vartheta >0$\ corresponds to the right rotation of the light
polarization plane and $\vartheta <0$\ corresponds to the left one, where
the right (positive) rotation is recording by the light observer as the
clockwise one.

In accordance with (\ref{6}) the T-odd interaction results in the photon
polarization plane rotation around the electric field $\overrightarrow{E}$%
{\LARGE \ }direction. The angle of rotation is proportional to the
polarizability $\beta _{E}^{T}$ and the $(\overrightarrow{n}_{E}%
\overrightarrow{n}_{\gamma })$ correlation. Together with the T-odd effect
there is \ the\ well knownT-even P-odd polarization plane rotation
phenomenon \cite{7,10,11} determining by the polarizability $\beta _{s}^{P}$ and
being independent on the $(\overrightarrow{n}_{E}\overrightarrow{n}_{\gamma
})$ correlation. The T-odd rotation dependence on the electric field {\LARGE %
\ }$\overrightarrow{E}$ orientation with respect to the $\overrightarrow{n}%
_{\gamma }$ direction allows one to distinguish T-odd and T-even P-odd
phenomena experimentally.

The refractive index $N_{+}(N_{-})$\ has both real and imaginary parts. The
imaginary part of the refractive index $({Im}N_{\pm }\sim {Im}\beta _{E}^{T}(%
\overrightarrow{n}_{E}\overrightarrow{n}_{\gamma }))$\ is responsible for
the T-reversal violating circular dichroism. Due to this process the
linearly polarized photon takes circular polarization. The sign of the
circular polarization depends on the sign of the scalar production $(%
\overrightarrow{n}_{E}\overrightarrow{n}_{\gamma })$\ that allows us to
separate T-odd circular dichroism from P-odd T-even circular dichroism. The
last one is proportional to ${Im}\beta _{s}^{P}$.

\section{\protect\Large T, P-odd polarisabilities of atoms and molecules.}

In order to estimate the magnitude of the effects one should obtain the
T-odd polarizability tensor $\alpha _{ik}^{odd}$\ and T-odd polarizabilities
\ of atom (molecule) or (that is actually the same, see (\ref{amplitude_ik},\ref{6})) 
the T-odd part of the amplitude $f(0)$ of elastic coherent scattering of
a photon by an atom (molecule).

According to quantum electrodynamics the elastic coherent
scattering at zero angle can be considered as the succession of two
processes: the first one is the absorption of the initial photon with the
momentum $\overrightarrow{k}$ and the transition of the atom (molecule) from
the initial state $\left| N_{0}\right\rangle $ with the energy $E_{N_{0}}$
to an intermediate state $\left| F\right\rangle $ with an energy $E_{F}$;
the second one is the transition of the atom (molecule) from the state $%
\left| F\right\rangle $ to the final state $\left| F^{\prime }\right\rangle
=\left| N_{0}\right\rangle $ and irradiation of the photon with the momentum 
$\ \overrightarrow{k}^{\prime }=\overrightarrow{k}$.

Let $H_{A}$\ be the atom (molecule) Hamiltonian considering the weak
interaction between electrons and nucleus and the electromagnetic
interaction of an atom (molecule) with the external electric $%
\overrightarrow{E}$ and magnetic $\overrightarrow{H}$ fields. It defines the
system of eigenfunctions $\left| F\right\rangle $\ and eigenvalues $%
E_{F}=E_{F}(\overrightarrow{E,}\overrightarrow{H})$: 
\begin{equation}
H_{A}\left| F\right\rangle =E_{F}\left| F\right\rangle ,  \label{7}
\end{equation}
$F-$set of quantum numbers describing the state $\left| F\right\rangle $.

According to \cite{PLA} T-odd effects in a gas exposed to external electric
and magnetic fields are manifested even in electric dipole approximation in
contrast to P-odd T-even phenomenon of light polarization plane rotation.
Polarizability correspondent to the latter one is proportional to product of
matrix elements of electric dipole and magnetic dipole transitions.

The matrix element of the process determining the scattering amplitude in
the forward direction in the dipole approximation is given by \cite{8}: 
\begin{equation}
{\frak M}_{N_{0}}=\sum_{F}\left\{ \frac{\left\langle N_{0}\right| 
\overrightarrow{d}\overrightarrow{e}^{\ast }\left| F\right\rangle
\left\langle F\right| \overrightarrow{d}\overrightarrow{e}\left|
N_{0}\right\rangle }{E_{F}-E_{N_{0}}-\hbar \omega }+\frac{\left\langle
N_{0}\right| \overrightarrow{d}\overrightarrow{e}\left| F\right\rangle
\left\langle F\right| \overrightarrow{d}\overrightarrow{e}^{\ast }\left|
N_{0}\right\rangle }{E_{F}-E_{N_{0}}+\hbar \omega }\right\} ,  \label{8}
\end{equation}
where $\overrightarrow{d}$\ is the electric dipole transition operator, $%
\omega $\ is the photon frequency, $\left| N_{0}\right\rangle $ is the wave
function of the initial state of atom (molecule) with the energy $E_{N_{0}}$%
, $\left| F\right\rangle $ is the wave function of an intermediate state
with an energy $E_{F}$.

For gases the energy of atom $E_F$ is composed from the internal energy of atom,
its kinetic energy and energy of atom interaction with external fields. It
is very important to note that in addition to interaction of atom with
fields $\overrightarrow{E}$ and $\overrightarrow{H}$\ it undergoes influence
caused by coherent scattering of an atom by other atoms of gas. Potential
energy of this interaction \cite{4,5} is

\[
U_{F}=-\frac{4\pi \hbar ^{2}}{M_{A}}\rho \;f_{F}(\overrightarrow{\varkappa },%
\overrightarrow{\varkappa }), 
\]
where $\rho $\ is the density of atoms of gas, $M_A$ is the mass of atom, 
$f_{F}(\overrightarrow{%
\varkappa },\overrightarrow{\varkappa })$ is the amplitude of forward
elastic coherent scattering of the atom being in the state $\left|
F\right\rangle $\ by atom of gas (if atoms are identical then exchange
scattering also contributes in this amplitude). Energy $U_{F}$ depends on
the atom state $\left| F\right\rangle $\ and orientation of the total moment
of an atom in this state. Thus it is fundamentally important to consider $%
U_{F}$\ in detail calculation because it can contribute in the effects
discussed.

Motion of gas atoms yields to Doppler shift of levels and in order to get
the final expressions the equation (\ref{6}) should be averaged over atom
momenta distribution in gas.Hereinafter we will not draw explitly this
routine procedure \cite{7}   

It should be reminded that the atom (molecule) exited levels are
quasistationary i.e. $E_{F}$\ has the imaginary part\ and everywhere $E_{F}$
should be presented as $(E_{F}-\frac{i}{2}\Gamma _{F})$\ , where $E_{F}$\ is
the atom (molecule) level energy, \ $\Gamma _{F}$ is the level width. 

The
matrix element (\ref{8}) can be written as: 
\begin{equation}
{\frak M}_{N_{0}}=\alpha _{ik}^{N_{0}}e_{i}^{\ast }e_{k},  \label{9}
\end{equation}
where $\alpha _{ik}^{N_{0}}$ is the tensor of dynamical polarizability of an
atom (molecule)

\begin{equation}
\alpha _{ik}^{N_{0}}=\sum_{F}\left\{ \frac{\left\langle N_{0}\right|
d_{i}\left| F\right\rangle \left\langle F\right| d_{k}\left|
N_{0}\right\rangle }{E_{F}-E_{N_{0}}-\hbar \omega }+\frac{\left\langle
N_{0}\right| d_{k}\left| F\right\rangle \left\langle F\right| d_{i}\left|
N_{0}\right\rangle }{E_{F}-E_{N_{0}}+\hbar \omega }\right\}  \label{10}
\end{equation}
In general case atoms are distributed to the levels of ground state $N_{0}$\
with the probability $P(N_{0})$. Therefore, $\alpha _{ik}^{N_{0}}$\ should
be averaged over all states $N_{0}$. As a result, the polarizability can be
written

\begin{equation}
\alpha _{ik}=\sum_{N_{0}}P(N_{0})\alpha _{ik}^{N_{0}}  \label{10-1}
\end{equation}
The tensor $\alpha _{ik}$ can be expanded in the irreducible parts as 
\begin{equation}
\alpha _{ik}=\alpha _{0}\delta _{ik}+\alpha _{ik}^{s}+\alpha _{ik}^{a},
\label{11}
\end{equation}
where $\alpha _{0}=\frac{1}{3}{\sum_{i}\alpha _{ii}}$ is the scalar, $\alpha
_{ik}^{s}=\frac{1}{2}(\alpha _{ik}+\alpha _{ki})-$\ $\alpha _{0}\delta _{ik}$%
\ is the symmetric tensor of rank two, $\alpha _{ik}^{a}=\frac{1}{2}(\alpha
_{ik}-\alpha _{ki})$ is the antisymmetric tensor of rank two,

\begin{eqnarray}
\alpha _{0} &=&\frac{2}{3}\sum_{N_{0}}P(N_{0})\sum_{iF}\frac{\omega
_{N_{0}}\left\langle N_{0}\right| d_{i}\left| F\right\rangle \left\langle
F\right| d_{i}\left| N_{0}\right\rangle }{\hbar (\omega _{FN_{0}}^{2}-\omega
^{2})}  \nonumber \\
\alpha _{ik}^{s} &=&\sum_{N_{0}}P(N_{0})\sum_{F}\frac{\omega
_{N_{0}}[\left\langle N_{0}\right| d_{i}\left| F\right\rangle \left\langle
F\right| d_{k}\left| N_{0}\right\rangle +\left\langle N_{0}\right|
d_{k}\left| F\right\rangle \left\langle F\right| d_{i}\left|
N_{0}\right\rangle ]}{\hbar (\omega _{FN_{0}}^{2}-\omega ^{2})}-\alpha
_{0}\delta _{ik}  \\
\alpha _{ik}^{a} &=&\frac{\omega }{\hbar }\sum_{N_{0}}P(N_{0})\sum_{F}\frac{%
\left\langle N_{0}\right| d_{i}\left| F\right\rangle \left\langle F\right|
d_{k}\left| N_{0}\right\rangle -\left\langle N_{0}\right| d_{k}\left|
F\right\rangle \left\langle F\right| d_{i}\left| N_{0}\right\rangle }{\omega
_{FN_{0}}^{2}-\omega ^{2}}  \nonumber
\end{eqnarray}
where $\omega _{FN_{0}}=\frac{E_{F}-E_{N_{0}}}{\hbar }$.

Let atoms (molecules) be nonpolarized. The antisymmetric part of
polarizability $\alpha_{ik}^a $ 
is equal to zero in the absence of T- and P- odd
interactions. It should be reminded that according to the above for the
P-odd and T-even interactions the antisymmetric part of polarizability
differs from zero only while considering both the electric and magnetic dipole transitions 
\cite{7}.

As it was shown above, scalar, symmetric and antisymmetric parts of 
$\epsilon_{ik}$ (and, therefore,$\alpha _{ik}$) describes some new T- 
and P-odd effects.
For example, the effect of polarisation plane rotation is described
by the antisymmetric part \ $\alpha _{ik}^{a}$ of tensor of dynamical
polarizability of atom (molecule) $\alpha _{ik}$ . We can evaluate $\alpha
_{ik}^{a}$\ and, as a result, obtain the expression for $\beta _{E}^{T}$\ by
the following way. According to (4,6) the magnitude of the T-odd effect is
determined by the polarizability $\beta _{E}^{T}$\ or (that is actually the
same, see (\ref{6})) by the amplitude $f_{\pm }(0)$\ of elastic
coherent scattering of a photon by an atom (molecule). If $\overrightarrow{n}%
_{E}\parallel \overrightarrow{n}_{\gamma }$\ the amplitude $f_{\pm }(0)$\ in
the dipole approximation can be written as $f_{\pm }=\mp \frac{\omega ^{2}}{%
c^{2}}\beta _{E}^{T}$. As a result, in order to obtain the amplitude $f_{\pm
}$, the matrix element (\ref{8},\ref{9}) for photon polarization states $%
\overrightarrow{e}=\overrightarrow{e}_{\pm }$\ should be found.

The electric dipole transition operator$\overrightarrow{d}$\ can be written
in the form: 
\begin{equation}
\overrightarrow{d}=d_{+}\overrightarrow{e}_{+}+d_{-}\overrightarrow{e}%
_{-}+d_{z}\overrightarrow{n}_{\gamma }\text{,}  \label{13}
\end{equation}
with $\overrightarrow{d}_{+}=-\frac{d_{x}-id_{y}}{\sqrt{2}}$, $%
\overrightarrow{d}_{-}=\frac{d_{x}+id_{y}}{\sqrt{2}}$. Let photon
polarization state $\overrightarrow{e}=\overrightarrow{e}_{+}$. Using (\ref
{8},\ref{9}) we can present the polarizability $\beta _{E}^{T}$\ as follows: 
\begin{equation}
\beta _{E}^{T}=\frac{\omega }{\hbar }\sum_{N_{0}}P(N_{0})\sum_{F}\left\{ 
\frac{\left\langle N_{0}\right| d_{-}\left| F\right\rangle \left\langle
F\right| d_{+}\left| N_{0}\right\rangle -\left\langle N_{0}\right|
d_{+}\left| F\right\rangle \left\langle F\right| d_{-}\left|
N_{0}\right\rangle }{\omega _{FN_{0}}^{2}-\omega ^{2}}\right\} .  \label{14}
\end{equation}

For further analysis the more detailed expressions for atom (molecule) wave
functions are necessary. The weak interaction constant is very small.
Therefore, we can use the perturbation theory. Let $\left| f,E\right\rangle $
be the wave function of an atom (molecule) interacting with an electric
field $\overrightarrow{E}$\ in the absence of weak interaction . Switch on
weak interaction $(V_{w}\neq 0)$. According to the perturbation theory\ \cite
{8}\ the wave function $\left| F\right\rangle $\ can be written in this case
as: 
\begin{equation}
\left| F\right\rangle =\left| f,\overrightarrow{E}\right\rangle +\sum_{n}%
\frac{\left\langle n,\overrightarrow{E}\right| V_{w}\left| f,\overrightarrow{%
E}\right\rangle }{E_{f}-E_{n}}\left| n,\overrightarrow{E}\right\rangle
\label{15}
\end{equation}
It should be mentioned that both numerator and denominator of (\ref{14})
contain $V_{w}$. Suppose $V_{w}$\ to be small one can represent the total
polarizability $\beta _{E}^{T}$\ as the sum of two terms

\begin{equation}
\beta _{E}^{T}=\beta _{mix}^{T}+\beta _{split}^{T},  \label{sum}
\end{equation}
where 
\begin{equation}
\beta _{mix}^{T}=\frac{\omega }{\hbar }\sum_{N_{0}}P(N_{0})\sum_{f}\frac{1}{%
\omega _{fn_{0}}^{2}-\omega ^{2}}\sum_{l}  \label{16}
\end{equation}

\bigskip $\left\{ \frac{2 {Re}\left[ \left\{ \left\langle n_{0}%
\overrightarrow{E}\right| d_{-}\left| f\overrightarrow{E}\right\rangle
\left\langle f\overrightarrow{E}\right| d_{+}\left| l\overrightarrow{E}%
\right\rangle -\left\langle n_{0}\overrightarrow{E}\right| d_{+}\left| f%
\overrightarrow{E}\right\rangle \left\langle f\overrightarrow{E}\right|
d_{-}\left| l\overrightarrow{E}\right\rangle \right\} \left\langle l%
\overrightarrow{E}\right| V_{w}\left| n_{0}\overrightarrow{E}\right\rangle %
\right] }{E_{n_{0}}-E_{l}}\right. +$

\bigskip$\left. +\frac{2 {Re}\left[ \left\langle n_{0}\overrightarrow{E}%
\right| d_{-}\left| l\overrightarrow{E}\right\rangle \left\langle l%
\overrightarrow{E}\right| V_{w}\left| f\overrightarrow{E}\right\rangle
\left\langle f\overrightarrow{E}\right| d_{+}\left| n_{0}\overrightarrow{E}%
\right\rangle -\left\langle n_{0}\overrightarrow{E}\right| d_{+}\left| l%
\overrightarrow{E}\right\rangle \left\langle l\overrightarrow{E}\right|
V_{w}\left| f\overrightarrow{E}\right\rangle \left\langle f\overrightarrow{E}%
\right| d_{-}\left| n_{0}\overrightarrow{E}\right\rangle \right] }{%
E_{f}-E_{l}}\right\} $

\noindent and

\begin{eqnarray}
\beta _{split}^{T} &=&\frac{\omega }{\hbar }\sum_{N_{0}}P(N_{0})\sum_{F}%
\left\{ \frac{\left\langle n_{0}\right| d_{-}\left| f\right\rangle
\left\langle f\right| d_{+}\left| n_{0}\right\rangle -\left\langle
n_{0}\right| d_{+}\left| f\right\rangle \left\langle f\right| d_{-}\left|
n_{0}\right\rangle }{\omega _{FN_{0}}^{2}-\omega ^{2}}\right\} =  \nonumber
\\
&=&\frac{\omega }{\hbar }\sum_{N_{0}}P(N_{0})\sum_{F}\left\{ \frac{%
\left\langle n_{0}\right| d_{-}\left| f\right\rangle \left\langle f\right|
d_{+}\left| n_{0}\right\rangle -\left\langle n_{0}\right| d_{+}\left|
f\right\rangle \left\langle f\right| d_{-}\left| n_{0}\right\rangle }{%
(\omega _{FN_{0}}-\omega )(\omega _{FN_{0}}+\omega )}\right\} =  \nonumber \\
&=&\frac{1}{2\hbar }\sum_{N_{0}}P(N_{0})\sum_{F}\left\{ \frac{\left\langle
n_{0}\right| d_{-}\left| f\right\rangle \left\langle f\right| d_{+}\left|
n_{0}\right\rangle -\left\langle n_{0}\right| d_{+}\left| f\right\rangle
\left\langle f\right| d_{-}\left| n_{0}\right\rangle }{(\omega
_{FN_{0}}-\omega )}\right\}  \label{16-1}
\end{eqnarray}
\[
\omega _{FN_{0}}=\frac{E_{F}(\overrightarrow{E})-E_{N_{0}}(\overrightarrow{E}%
)}{\hbar }, 
\]
It should be reminded that according to all the above (see also section 3)
energy levels $E_{F}$ and $E_{N_{0}}$ contain shifts caused by interaction
of electric dipole moment of the level with electric field $\overrightarrow{E%
}$ and magnetic moment of the level with T-odd induced magnetic field $%
\overrightarrow{H}_{ind}(\overrightarrow{E})$.

It should be noted that radial parts of the atom wave functions are real 
\cite{8,9}, therefore the matrix elements of operators $d_{\pm }$ are real
too. As a result, the P-odd T-even part of the interaction $V_{w}$\ does not
contribute to $\beta _{mix}^{T}$\ \ because the P-odd T-even matrix elements
of $V_{w}$\ are imaginary \cite{7}.\ At the same time, the T- and P-odd
matrix elements of $V_{w}$\ are\ real \cite{7}, therefore, the
polarizability $\beta _{mix}^{T}\neq 0$. It should be mentioned that in the
absence of electric field ($\overrightarrow{E}=0$) the polarizability $\beta
_{E}^{T}=0$\ \ and, therefore, the phenomenon of the photon polarization
plane rotation is absent.

Really, the electric field $\overrightarrow{E}$\ mixes the opposite parity
levels of the atom . The atom levels have the fixed parity at $%
\overrightarrow{E}=0$. The operators\ $d_{\pm }$\ and $V_{w}$\ change the
parity of the atom states. As a result, the parity of the final state $%
\left| N_{0}^{\prime }\right\rangle =$\ \ $d_{+}$\ $d_{-}$\ $V_{w}$\ $\left|
N_{0}\right\rangle $\ appears to be opposite to the parity of the initial
state $\left| N_{0}\right\rangle $. But the initial and final states \ in
the expression for $\beta _{E}^{T}$\ are the same. Therefore $\beta _{E}^{T}$%
\ can not differ from zero\ at $\overrightarrow{E}=0$.

It should be emphasized once again that polarizability $\beta _{E}^{T}$\ \
differs from zero even if EDM of electron is equal to zero. The interaction
of electron EDM with electric field gives only part of contribution to the
total polarizability of atom (molecule). The new effect we discuss is caused
by the Stark mixing of atom (molecule) levels and weak T- and P-odd
interaction of electrons with nucleus (and with each other).

Therefore, according to (\ref{sum}) the total angle of polarization plane
rotation includes two terms $\vartheta =\vartheta _{mix}+\vartheta _{split}$%
, where \ $\vartheta _{mix}\sim \beta _{mix}^{T}$\ is caused by the
considered above effect similar to Van Vleck that and $\vartheta
_{split}\sim \beta _{split}^{T}$\ is caused by the \ atom levels splitting
both in electric field $\overrightarrow{E}$\ and magnetic field $%
\overrightarrow{H}_{ind}(\overrightarrow{E})$. The contributions given by $%
\beta _{mix}^{T}$\ and \ $\beta _{split}^{T}$\ can be distinguished by
studying the frequency dependence of $\vartheta =\vartheta (\omega )$.
According to (\ref{16},\ref{16-1}) $\vartheta _{mix}\sim \frac{1}{\omega
_{fn_{0}}-\omega }$\ whereas $\vartheta _{split}\sim \frac{1}{(\omega
_{fn_{0}}-\omega )^{2}}.$\ So, $\vartheta _{split}$\ \ decreases faster then 
$\vartheta _{mix}$ with the grows of frequency tuning out from resonance.

Let us now estimate the magnitude of the effect of the T-odd photon plane
rotation due to $\beta _{mix}^{T}$. According to the analysis \cite{4,5,6},
based on the calculations \ of the value of \ T- and P-noninvariant
interactions given by \cite{7}, the ratio $\frac{\left\langle
V_{w}^{T}\right\rangle }{\left\langle V_{w}^{P}\right\rangle }\leq
10^{-3}\div 10^{-4}$, where\ $\left\langle V_{w}^{T}\right\rangle $\ is T
and P-odd matrix element, $\left\langle V_{w}^{P}\right\rangle $\ is P-odd
T-even matrix element.

The P-odd T-even polarizability $\beta _{s}^{P}$\ is proportional to the product of
electric and magnetic dipole matrix elements and $%
\left\langle V_{w}^{P}\right\rangle $: $\beta _{s}^{P}\sim $\ $\left\langle
d\right\rangle \left\langle \mu \right\rangle \left\langle
V_{w}^{P}\right\rangle $\ \cite{7}. At the same time $\beta _{mix}^{T}\sim $%
\ $\left\langle d(\overrightarrow{E})\right\rangle \left\langle d(%
\overrightarrow{E})\right\rangle \left\langle V_{w}^{T}\right\rangle $. As a
result, 
\begin{equation}
\frac{\beta _{mix}^{T}}{\beta _{s}^{P}}\sim \frac{\left\langle d(%
\overrightarrow{E})\right\rangle \left\langle d(\overrightarrow{E}%
)\right\rangle \left\langle V_{w}^{T}\right\rangle }{\left\langle
d\right\rangle \left\langle \mu \right\rangle \left\langle
V_{w}^{P}\right\rangle }.  \label{17}
\end{equation}
Let us study the T-odd phenomena of the photon polarization plane rotation
in an electric field $\overrightarrow{E}$\ for the transition $%
n_{0}\rightarrow f$ \ between the levels $n_{0}$ and $f$\ which have the
same parity at $\overrightarrow{E}=0$.{\bf \ }The matrix element\ $%
\left\langle n_{0},\overrightarrow{E}\right| d_{\pm }\left| f,%
\overrightarrow{E}\right\rangle $\ does not equal to zero only if $%
\overrightarrow{E}\neq 0$ . Let the energy of interaction of an atom with an
electric field, $V_{E}=-$ $\overrightarrow{d}\overrightarrow{E},$\ be much
smaller than the spacing $\Delta $\ of the energy levels, which are mixed by
the field $\overrightarrow{E}$. Then one can use the perturbation theory for
the wave functions $\left| f,\overrightarrow{E}\right\rangle $: 
\begin{equation}
\left| f,\overrightarrow{E}\right\rangle =\left| f\right\rangle +\sum_{m}%
\frac{\left\langle m\right| -d_{z}E\left| f\right\rangle }{E_{f}-E_{m}}%
\left| m\right\rangle ,  \label{18}
\end{equation}
where $z\parallel \overrightarrow{E}$. As a result, the matrix element $%
\left\langle n_{0},\overrightarrow{E}\right| d_{\pm }\left| f,%
\overrightarrow{E}\right\rangle $\ \ can be rewritten as: 
\begin{eqnarray}
\left\langle n_{0},\overrightarrow{E}\right| d_{\pm }\left| f,%
\overrightarrow{E}\right\rangle &=&  \label{19} \\
&=&-\left\{ \sum_{m}\frac{\left\langle n_{0}\right| d_{\pm }\left|
m\right\rangle \left\langle m\right| d_{z}\left| f\right\rangle }{E_{f}-E_{m}%
}+\right.  \nonumber \\
&&+\left. \sum_{p}\frac{\left\langle n_{0}\right| d_{z}\left| p\right\rangle
\left\langle p\right| d_{\pm }\left| f\right\rangle }{E_{n_{0}}-E_{p}}%
\right\} E.  \nonumber
\end{eqnarray}
One can see that the matrix element $\left\langle d(\overrightarrow{E}%
)\right\rangle \sim \ \frac{\left\langle d\right\rangle E}{\Delta }$\ $%
\left\langle d\right\rangle $\ in this case. The other matrix elements in (%
\ref{16}) can be evaluated at $\overrightarrow{E}=0$. This gives the
estimate as follows: 
\begin{equation}
\beta _{mix}^{T}\sim \left\langle d\right\rangle \left\langle d\right\rangle 
\frac{\left\langle dE\right\rangle }{\Delta }\left\langle
V_{w}^{T}\right\rangle \ \ .  \label{20}
\end{equation}
and, consequently, ratio (\ref{17}) can be written as 
\begin{equation}
\frac{\beta _{mix}^{T}}{\beta _{s}^{P}}\sim \frac{\left\langle
d\right\rangle \left\langle d\right\rangle \frac{\left\langle
dE\right\rangle }{\Delta }\left\langle V_{w}^{T}\right\rangle }{\left\langle
d\right\rangle \left\langle \mu \right\rangle \left\langle
V_{w}^{P}\right\rangle }.  \label{21}
\end{equation}
Taking into account that the matrix element $\left\langle \mu \right\rangle
\sim \alpha \left\langle d\right\rangle $\ \ \cite{8,9}, where $\alpha =%
\frac{1}{137}$ is the fine structure constant, equation (\ref{21}) can be
reduced to: 
\begin{equation}
\frac{\beta _{mix}^{T}}{\beta _{s}^{P}}\sim \alpha ^{-1}\frac{\left\langle
dE\right\rangle }{\Delta }\frac{\left\langle V_{w}^{T}\right\rangle }{%
\left\langle V_{w}^{P}\right\rangle }  \label{22}
\end{equation}
For the case $\frac{\left\langle dE\right\rangle }{\Delta }\sim 1$, ratio (%
\ref{22}) gives 
\begin{equation}
\frac{\beta _{mix}^{T}}{\beta _{s}^{P}}\sim \alpha ^{-1}\frac{\left\langle
V_{w}^{T}\right\rangle }{\left\langle V_{w}^{P}\right\rangle }\lesssim
10^{-1}\div 10^{-2}  \label{23}
\end{equation}
Such condition can be realized, for example, for exited states of atoms and
for two-atom molecules (TlF, BiS, HgF) which have a pair of nearly
degenerate opposite parity states. As one can see, the ratio $\frac{\beta
_{mix}^{T}}{\beta _{s}^{P}}$\ is two orders larger as compared with the
simple estimation $\frac{\left\langle V_{w}^{T}\right\rangle }{\left\langle
V_{w}^{P}\right\rangle }\leq 10^{-3}\div 10^{-4}$\ due to the fact that $%
\beta _{mix}^{T}$\ is determined by only the electric dipole transitions ,
while $\beta _{s}^{P}$\ is determined by both the electric and magnetic
dipole transitions.

\section{{\bf {\protect\Large The possibility to observe the time-reversal violating optical 
phenomena
experimentally.}}}

The possibility to observe the time-reversal violating optical phenomena
experimentally can be discussed
now. In accordance with (6) the angle of the T-odd rotation in electric
field can be evaluated as follows: 
\begin{equation}
\vartheta _{mix}^{T}\sim \frac{2\pi \rho \omega }{c}\beta _{mix}^{T}L\sim 
\frac{\beta _{mix}^{T}}{\beta _{S}^{P}}\vartheta ^{P}\sim \alpha ^{-1}\frac{%
\left\langle dE\right\rangle }{\Delta }\frac{\left\langle
V_{w}^{T}\right\rangle }{\left\langle V_{w}^{P}\right\rangle }\vartheta ^{P}.
\label{2-22}
\end{equation}
According to the experimental data \cite{10,11} being well consistent with
calculations \cite{7} the typical value of $\vartheta ^{P}$\ is $\vartheta
^{P}\sim 10^{-6}rad$\ (for the length $L$\ being equal to the several
absorption lengths of the light propagating through a gas $L_{a}$).

For the electric field $E\sim 10^{4}V\cdot cm^{-1}$\ the parameter $\frac{%
\left\langle dE\right\rangle }{\Delta }$\ can be estimated as $\frac{%
\left\langle dE\right\rangle }{\Delta }\sim 10^{-5}$\ for Cs, Tl and $\frac{%
\left\langle dE\right\rangle }{\Delta }\sim 10^{-4}$\ for Yb and lead.
Therefore, one can obtain $\vartheta _{mix}^{T}\sim 10^{-13}rad$\ for Cs, Tl
and $\vartheta _{mix}^{T}\sim 10^{-12}rad$\ for Yb and lead. For the
two-atom molecules (TlF, BiS, HgF) the angle $\vartheta _{mix}^{T}$\ can be
larger, because they have a pair of degenerate opposite parity states.

It should be noted that the classical up-to-date experimental techniques
allow to measure angles of light polarization plane rotation up to $4,3\cdot
10^{-11}rad$\ \ \cite{tarasov}.

A way to increase the rotation angle $\vartheta ^{T}$\ is to increase the
length $L$\ \ of the path of a photon inside a medium (see (6)). It can be
done, for example, by placing a medium in a resonator or inside a laser
gyroscope. This becomes possible due to the fact that in
contrast with the phenomenon of P-odd rotation of the polarization plane of
photon the T-odd rotation in an electric field is accumulated while photon
is moving both in the forward and backward directions. 

\begin{figure}[htbp]
\epsfxsize = 13,5 cm \centerline{\epsfbox{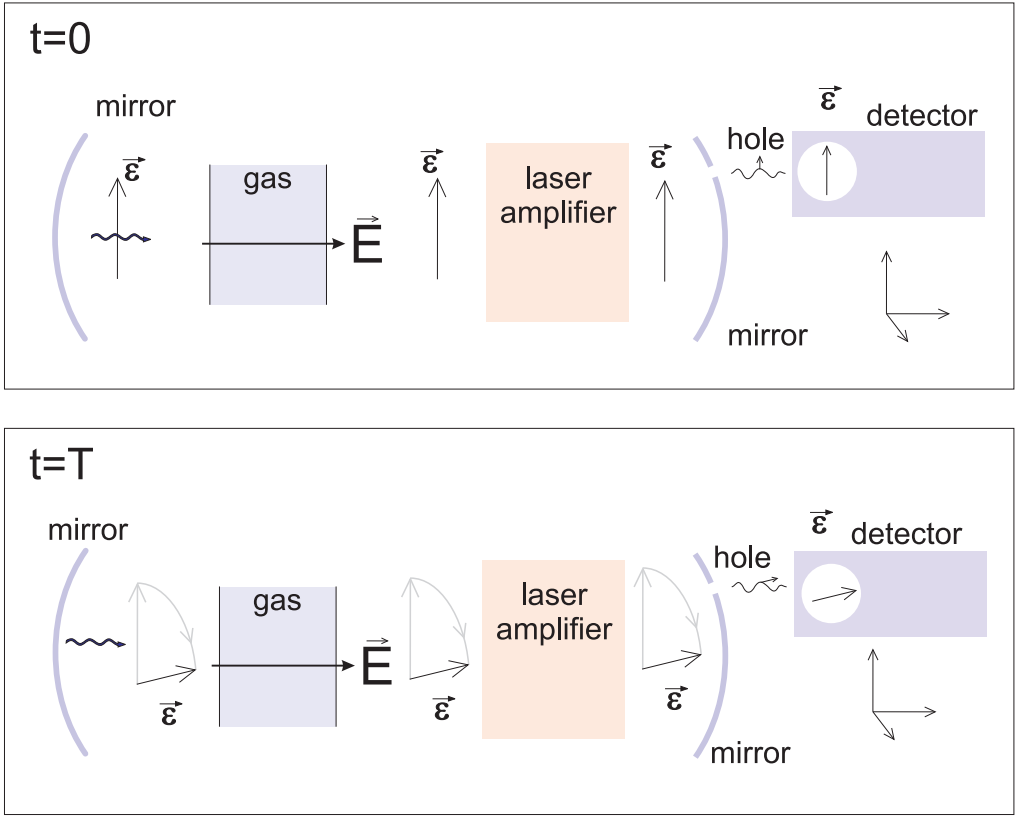}}
\caption{Polarization plane of photons in a trap rotates in time around the electric field direction.}
\label{F1}
\end{figure}

For the first view the re-reflection of the wave in resonator can not
provide the significant increase of the photon path length $L$\ in
comparison with the absorption length $L_{a}$\ because of the absorption of
photons in a medium. Nevertheless this difficulty can be overcome when the
part of resonator is filled by the amplifying medium (for example, inverse
medium). As a result, the electromagnetic wave being absorbed by the
investigated gas is coherently amplified in the amplifier and then is
refracted to the gas again. Consequently, under the ideal conditions the
light pulse can exist in such resonator-amplifier for arbitrarily long time.
And the peculiar ''photon trap'', in which photon polarisation plane
rotates, appears (Fig.\ref{F1}). The angle of rotation $\vartheta _{t}^{T}=\Omega ^{T}\cdot
t$, where $\Omega ^{T}$\ is the frequency of the photon polarization plane
rotation around the $\overrightarrow{E}$\ \ direction, $t$\ is the time of
electromagnetic wave being in a ''trap''. It is easy to find the frequency $%
\Omega ^{T}$\ from (6): $\ \Omega ^{T}=\frac{\vartheta ^{T}}{L}c=2\pi \rho
\omega \beta _{E}^{T}$. From the estimates of $\vartheta ^{T}$\ it is
evident that for $\vartheta ^{T}\sim 10^{-12}\;rad$\ (Lead, Yb) \ the
frequency $\Omega ^{T}$\ appears to be \ $\Omega ^{T}=\frac{\vartheta ^{T}}{%
L_{a}}c\sim 10^{-4}\sec ^{-1}$. Therefore $\vartheta _{t}^{T}\sim 10^{-4}t$\
and for the time $t$ of about 3 hours the angle $\vartheta _{t}^{T}$\
becomes $\sim 1\;rad$. The similar estimates for the atoms Cs, Tl ($%
\vartheta ^{T}\sim 10^{-13}\;rad$) give that for the same time the angle $%
\vartheta _{t}^{T}\sim 10^{-1}\;rad$.

The time $t$\ is limited, in particular, by spontaneous radiation of photons
in an amplifier that gradually leads to the depolarization of photon gas in
resonator. Surely, it is the ideal picture, but here is the way to further
increase of the experiment sensitivity. The achieved sensitivity in
measurements of phase incursion in laser gyroscope makes possible to observe
the effect in laser gyroscope, too. Laser interferometers used as
gravitational wave detectors also can provide neccessary sensitivity.

Requiring to measure rotation angle $\sim $10$^{-6}{}rad$\ in ''photon
trap'' and taking into consideration that existing technique allows to
measure much less angles one can expect to observe effect of the order $%
\frac{V_{T}}{V_{P}}\sim 10^{-9}\div 10^{-10}.$

All the estimations discussed above for the photon polarisation plane
rotation phenomena can be fully refered to birefringence effect which is
described by the symmetrical part of T-odd polarisability tensor. Two
effects contribute both in polarisation plane rotation and birefringence
phenomena. They are:    

1. splitting of atom (molecule) levels

2. mixing of ground state and opposite parity states in external fields

Birefringence effect appears when photon moves orthogonally to the electric
and magnetic fields. The magnitude of effect is proportional to the
correlation \ $(\overrightarrow{E}\overrightarrow{H})$.\ Thus, one can
distinguish the T-odd birefringence effect against a background of T-even
one by changing the direction of $\overrightarrow{E}$ with respect to
direction of $\overrightarrow{H}$. 

\begin{figure}[htbp]
\epsfxsize = 13,5 cm \centerline{\epsfbox{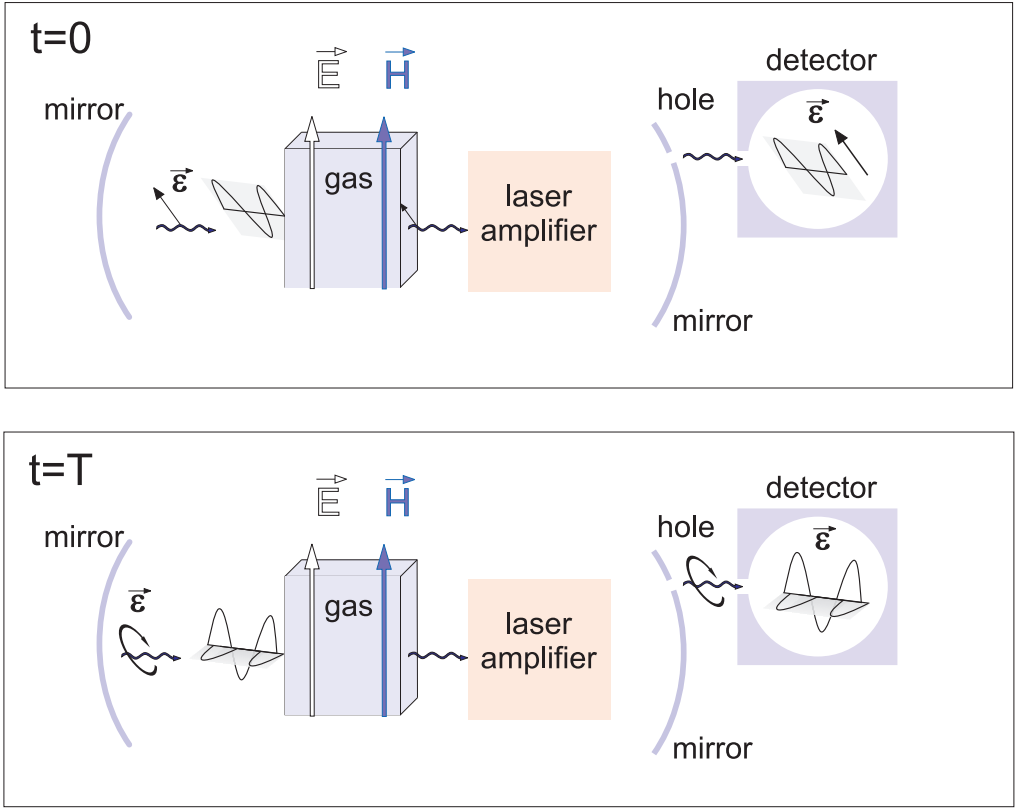}}
\caption{Owing to birefringence effect in ''photon trap'' linear  polarisation is converted
into circular one and vice versa}
\label{F2}
\end{figure}

For birefringence effect it should be mentioned that change\ of polarisation
type occurs in ''photon trap'' in time: circular polarisation is converted
into linear one, then   linear polarisation is converted into circular one
and so on (see Fig. (\ref{F2}))

Suppose effects of polarization plane rotation and birefringence be caused
only by atom EDM one can estimate the possible sensitivity of EDM
measurement in such experiments. Suppose we will measure rotation angle with
sensitivity about 10$^{-6}{}rad/hour$ (degree of circular polarisation in
the birefringence effect $\delta \sim$10$^{-6}{}\;per/hour$).\
Rotation angle is 
$\delta \vartheta =k\; Re(N_{+}-N_{-})L=k\; Re(N_{+}-N_{-})cT$, 
where $T$ is the observation
time (degree of circular polarisation at convertation from linear
polarisation to circular that 
$\delta =k\; Re(N_{_{\parallel }}-N_{_{\perp}})L=k\; Re(N_{_{\parallel }}-N_{_{\perp }})cT$ ), here 
$N_{_{\parallel }}$ is the index of refraction of photon with linear polarization parallel to the electric field and
$N_{_{\perp}}$ is that perpendicular to the electric field. 
Representing $\delta \vartheta $\ in the form \ \ 
\begin{equation}
\delta \vartheta =\frac{\rho cT\lambda ^{2}}{2\pi }\frac{\Gamma _{e}d_{a}E}{%
\hbar \Gamma ^{2}}  \label{delta fi for EDM}
\end{equation}
where $\rho $\ is the atoms density,\ $\Gamma _{e}$\ is the level radiation
width, $\Gamma $\ is the atom level width (including Doppler widening), \ $E$%
\ is the electric field strength one can estimate $d_{a}$\ as 
\begin{equation}
d_{a}=\frac{2\pi \hbar \Gamma ^{2}}{\rho cT\lambda ^{2}\Gamma _{e}E}\delta
\vartheta \thickapprox 10^{-33}e  \label{EDM estimation}
\end{equation}
(here $\lambda \sim 10^{-4}cm$, $E=10^{2}CGSE$, $\rho =10^{17}$ $atoms/cm^{3}
$, $\delta \vartheta \thickapprox 10^{-6}$, $T=1\;hour=3,6\cdot 10^3\;sec$). 
The similar estimations can be achieved at analysis of birefringence
effect.

For comparison it is interesting to note that the best expected EDM
measurement limit in recent publications \cite{Budker} is about $\
d_{a}\thickapprox 10^{-28}e$\ , so the advantages of the proposed method
becomes evident.

All the said can be applied not only for the optical range but for the radio
frequency range as well where the observation of the mentioned phenomenon is
also possible by the use of the mentioned atoms and molecules \cite{6}.

Thus, we have shown that the T-odd and P-odd phenomena of photon
polarization plane rotation and circular dichroism in an electric field are
expected to be observable experimentally. 

\section{Conclusion}

Analysis of the tensor of dielectric permittivity of medium $\epsilon_{ik}$ in 
presence of P- and T-odd interactions results in conclusion of existence of
several interesting T-noninvariant optical gyrotropy phenomena. Despite they 
are small,the use of photon "trap" allow to expect to make possible the 
experimental observation that let us obtain the important data on the value 
of T-odd weak interactions.

\end{document}